**Mössbauer and X-ray Photoelectron Spectroscopy Studies of $Fe_2BiMO_7$ (M = Sb, Ta) pyrochlore compounds synthesized by molten salts method.**

Jesús Alberto León Flores[a], José Luis Pérez Mazariego[a,*], Shirley Saraí Flores Morales[a], Roberto Hinojosa Nava[a], Paola Arévalo López[a], Raúl Escamilla Guerrero[b] and Raúl W. Gómez González[a].

[a] Facultad de Ciencias, Universidad Nacional Autónoma de México, Apartado Postal 70-399, México CD. MX., 04510, México

[b] Instituto de Investigaciones en Materiales, Universidad Nacional Autónoma de México, CD. MX., 04510, México.

*Corresponding author: Tel. (+5255) 56224849; e-mail: mazariego@ciencias.unam.mx
Facultad de Ciencias, Universidad Nacional Autónoma de México, Apartado Postal 70-399, México CD. MX., 04510, México

**ABSTRACT**

Polycrystalline samples of $Fe_2BiMO_7$ (M = Sb, Ta) compounds with pyrochlore-type structure were synthesized for the first time by the molten salts method. The compounds were obtained in one hour at 950ºC. The structures were determined by Rietveld refinement. Through Mössbauer spectroscopy and X-ray photoelectron spectroscopy, the site occupancies and ionic states of the cations in the pyrochlore structure were investigated.

## 1. INTRODUCTION

Synthesis by the molten salts method represents an alternative route to the typical solid-state reaction to obtain ceramic powders. The molten salts synthesis uses a low melting point, water-soluble salts as a reaction medium that enhances the mobility of the powder oxide reactants, promoting the rapid diffusion at relatively low temperatures. This fact allows achieving the formation of ceramic products in relatively short times respect to the solid-state reaction synthesis [**1, 2**].

Pyrochlore type structure compounds have been subject of interest in the latest years because they possess many types of interesting properties. Some pyrochlores, like $Yb_2Ti_2O_7$, have useful applications for immobilization of lanthanides and actinides from nuclear fuel reprocessing [**3**]. Also, their photocatalytic properties with applications on hydrogen production and wastewater decontamination have been investigated [**4**, **5**]. A typical pyrochlore compound has the general formula $A_2B_2O_7$ with a cubic structure belonging to the space group Fd-3m (No 227), where the A cation is eight-coordinated, and the B cation is six-coordinated [**6**]. ¿eight-fold coordination?

Several bismuth iron antimony pyrochlores have been reported in past years. However, previous attempts present hematite ($\alpha$-$Fe_2O_3$) and iron antimonate ($FeSbO_4$) as secondary phases, because these are inherent phases formed in the solid-state reaction. We introduce an alternative method of synthesis, which allows us a considerable time saving, with no intermediate reagent compound [**7**, **8**]. This paper presents the synthesis of $Fe_2BiMO_7$ (M = Sb, Ta) compounds with pyrochlore-type structure by the molten salts method and their characterization with X-ray powder diffraction together with cation site occupancies investigation via Mössbauer Spectroscopy and X-ray photoelectron spectroscopy (XPS).

## 2. MATERIALS AND METHODS

Polycrystalline samples of $Fe_2BiMO_7$ (M = Sb, Ta) were synthesized by the molten salt method. The precursors were $Fe_2O_3$ (Sigma-Aldrich >99%), $Ta_2O_5$ (Sigma-Aldrich 99.99%), $Sb_2O_5$ (Sigma-Aldrich 99.995%), $Bi_2O_3$ (Sigma-Aldrich 99.99%) and a mixture of salts NaCl-KCl (1:1) (Sigma-Aldrich 99.9%) in a 4:1 molar proportion respect to oxide powder reactants for M = Sb and 5:1 molar proportion for M = Ta. In the reaction, the salts were first mixed in an agate mortar until a fine homogeneous powder was attained. The same procedure was followed for the precursor reactants and then both powders were mixed and heated at 950 °C for 1 hour. The product obtained was washed and stirred in distilled deionized water for 3 hours to remove the salts. Then it was filtered in a Nalgene® system using a 0.22µm Millipore® filter.

A first rapid X-ray scan suggested an incomplete reaction (Section 3.1), so a final wash of the samples was done with HCl (J.T. Baker at 36%) diluted at 28% at 60ºC for two hours to eliminate the remaining iron oxide phase [**9**]. The X-ray diffraction patterns were measured at room temperature using a Siemens (D5000) diffractometer with Co $K_\alpha$ radiation and a Fe filter in 0.02° steps from 10° to 120°. The X-ray spectra were refined with the Rietveld program MAUD [**10**].

The room temperature Mössbauer spectra were recorder with a constant acceleration spectrometer, using thin absorbers (20 $mg/cm^2$) made with the obtained powders and using a 25 mCi $^{57}Co$ source in a rhodium matrix. All the Mössbauer spectra were fitted with the Recoil 1.05 software [**11**].

X-ray photoelectron spectroscopy (XPS) measurements were performed in an ultra-high vacuum (UHV) system Scanning XPS microprobe PHI 5000 Versa Probe II, with an Al $K_\alpha$ X-ray source (hν= 1486.6 eV), and an MCD analyzer. The surface of the samples was etched for 10 to 30 minutes with 2.0 kV $Ar^+$. The XPS spectra were obtained at 45° to the normal surface in the constant pass energy mode, $E_0$ = 100 eV (surface survey) and 10 eV (high-resolution narrow scan). Deconvolution adjustment for the XPS spectra was made by a least square process with the Spectral Data Processor software [**12**].

## 3. RESULTS

### 3.1 X-ray diffraction and Rietveld refinement

Figure 1 shows the X-ray diffraction patterns of $Fe_2BiSbO_7$ and $Fe_2BiTaO_7$ synthesized by the molten salt method, before (figures 1. a and 1. c) and after washed with HCl (figures 1. b and 1. d). As can be noted, the reflections related to α-$Fe_2O_3$ phase clearly diminished after the HCl washing prosses. According to Rietveld analysis results, 20% of α-$Fe_2O_3$ and a 15% of α-$Fe_2O_3$ were present in the pristine $Fe_2BiSbO_7$ and $Fe_2BiTaO_7$ samples (Table I).

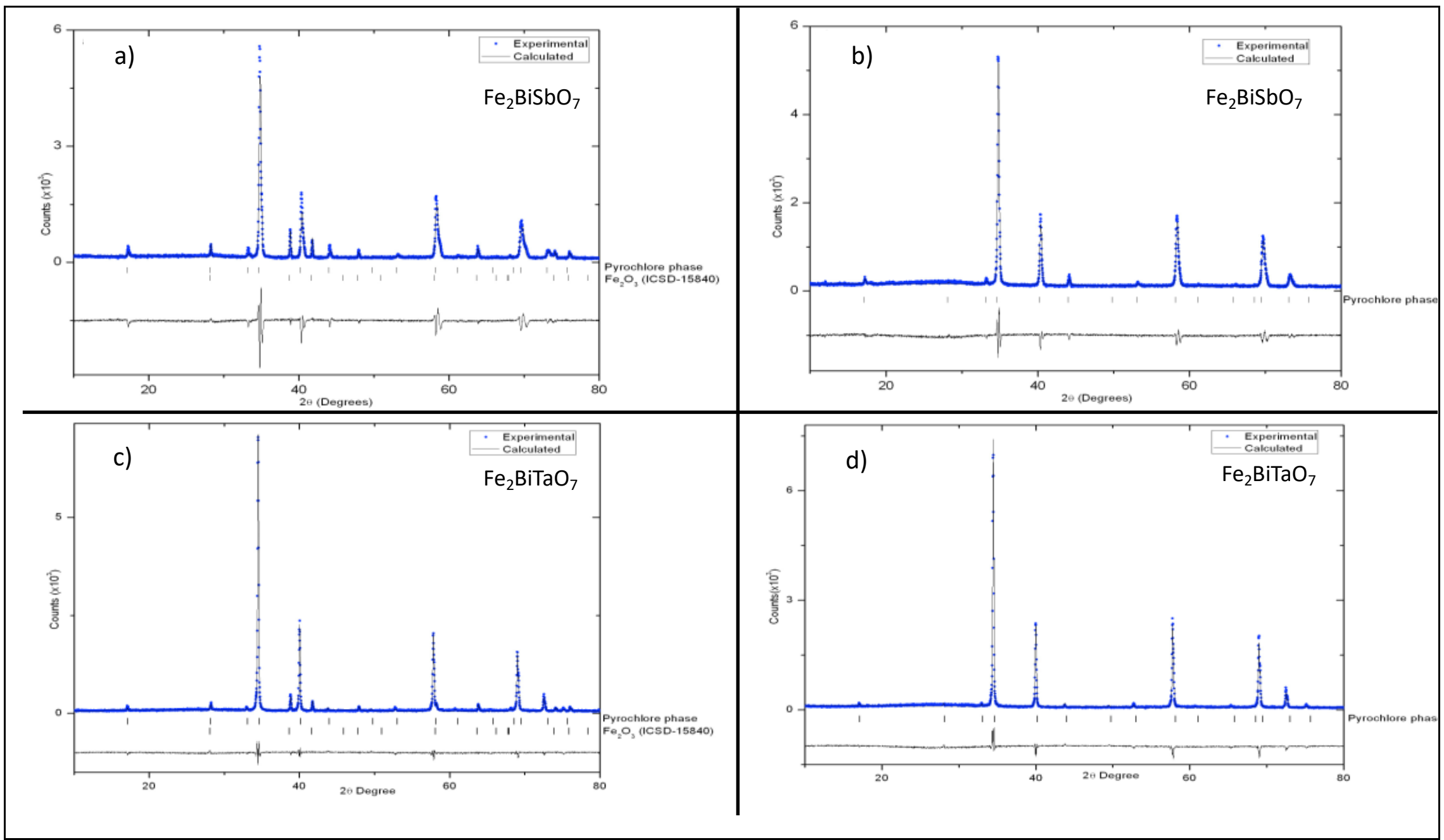

**Figure 1.** Rietveld analysis of the $Fe_2BiSbO_7$ and $Fe_2BiTaO_7$.

1a) and 1c) before washing with HCl. Pyrochlore phase and iron oxide (hematite) are observed

1b) and 1.d) after washing with HCl. Pure pyrochlore phase is observed.

For both synthesized samples, the A(16d) site occupancy was about 20%. In the $Fe_2BiSbO_7$ compound, iron cations occupy the A(16d) site, and in the $Fe_2BiTaO_7$ sample, the A(16d) site are occupied by the tantalum cations (table II), in good agreement with the peak intensity adjustment. These results match well with those previously reported [**7**, **8**]. It seems that the only effect of washing with HCl is to remove the unreacted $Fe_2O_3$ living a pure pyrochlore compound phase.

**Table I.** Rietveld refinement parameters of samples not washed with HCl and those who received an HCl treatment.

| Lattice parameter | $Fe_2BiSbO_7$ | $Fe_2BiSbO_7$ + HCl | $Fe_2BiTaO_7$ | $Fe_2BiTaO_7$ + HCl |
|---|---|---|---|---|
| a (Å) | 10.4077(5) | 10.4016(3) | 10.4927(1) | 10.4918(1) |
| % of phase | 80 | 100 | 85 | 100 |
| $Fe_2O_3$ | 20 | - | 15 | - |
| $\chi^2$ | 2.05 | 1.65 | 1.37 | 1.75 |

**Table II.** Structure parameters from the Rietveld refinement analysis for $Fe_2BiSbO_7$/$Fe_2BiTaO_7$ samples after HCl treatment.

| Site | x | y | z | Occupancy $Fe_2BiSbO_7$ | Occupancy $Fe_2BiTaO_7$ |
|---|---|---|---|---|---|
| A(16d) | 0.5 | 0.5 | 0.5 | 0.5 (Bi)<br>0.16 (Fe) | 0.50 (Bi)<br>0.20 (Ta) |
| B(16c) | 0 | 0 | 0 | 0.36 (Sb)<br>0.64 (Fe) | 0.30 (Ta)<br>0.70 (Fe) |
| O(48f) | 0.343(1)/0.329(9) | 0.125 | 0.125 | 1.00 | 1.00 |
| O(8b) | 0.375 | 0.375 | 0.375 | 1.00 | 1.00 |

| Bond length | $Fe_2BiSbO_7$ (Å) | $Fe_2BiTaO_7$ (Å) |
|---|---|---|
| A(16d)-O(48f) | 2.4567 | 2.5761 |
| A(16d)-O(8b) | 2.2521 | 2.2715 |
| B(16c)-O(48f) | 2.0796 | 2.0341 |

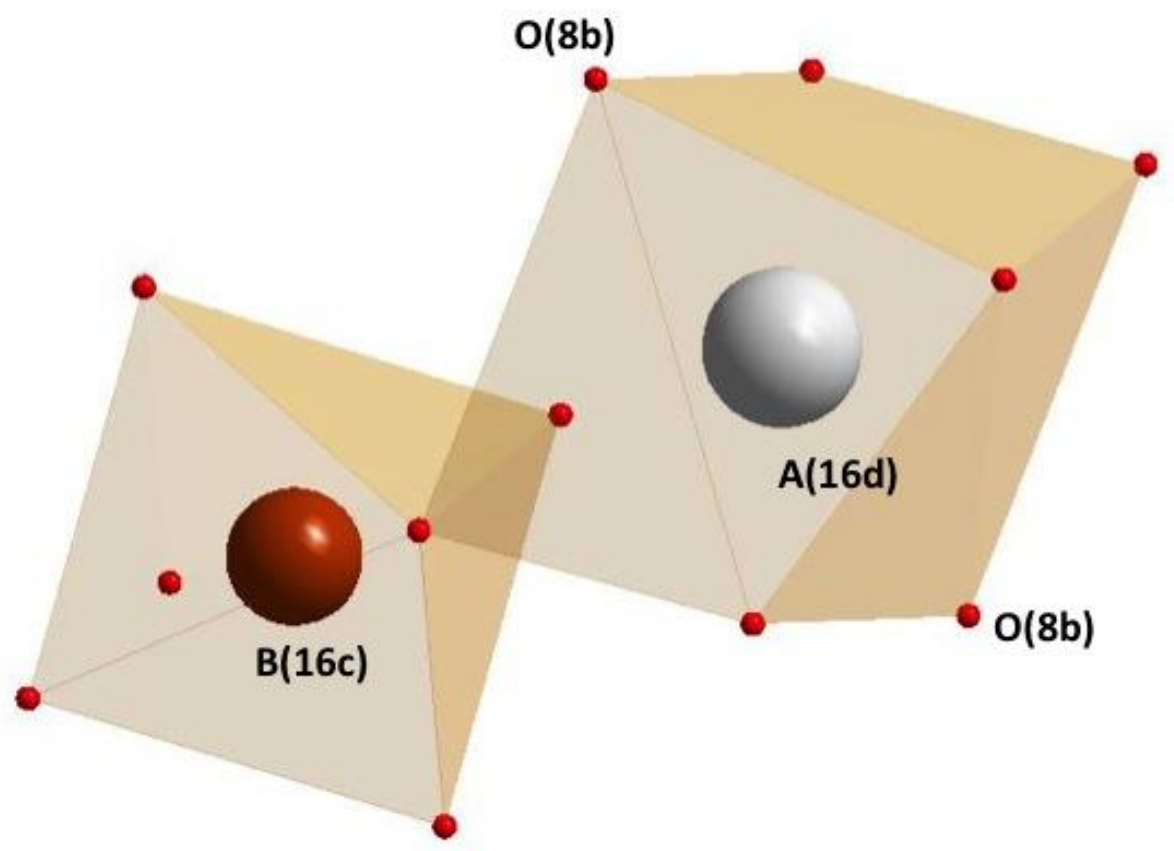

**Figure 2.** Crystal structure representation, from Rietveld refinement parameters, for the coordination around A(16d) and B(16c) cation sites.

## 3.2 Mössbauer spectroscopy

Figures 3. a) and 4. a) displays the Mössbauer spectra associated with the $Fe_2BiSbO_7$ and $Fe_2BiTaO_7$ samples, before HCl wash treatment. The Mössbauer spectrum of $Fe_2BiSbO_7$ consists of a sextet and two doublets, figure 3a). The sextet was fitted with an isomer shift ($\delta$) of 0.37 mm/s, with a quadrupole splitting ($\Delta Q$) of 0.10 mm/s and a magnetic field magnitude (H) of 51.41 T, corresponding to $\alpha$-$Fe_2O_3$, notwithstanding the 1% difference respect to reported value of 51.5 T [**13**], in accordance with the X-ray diffraction pattern that shows hematite phase. The presence of the two doublets in $Fe_2BiSbO_7$ sample leads us to conclude that iron cations occupy two non-equivalent positions inside the pyrochlore structure.

Doublet 1 was fitted with an $\delta$ = 0.34 mm/s and a $\Delta Q$ = 0.48 mm/s and a population of around 80%, associated with the B(16c) site while the doublet 2 was fitted with a $\delta$ = 0.34 mm/s, a $\Delta Q$ = 2.11 mm/s and an approximated population of 20%, related with the A(16d) site (table III).

The $Fe_2BiTaO_7$ sample, figure 4. a), also displays a sextet but only one doublet was observed. The sextet was fitted with $\delta$ = 0.36 mm/s and $\Delta Q$ = 0.1 mm/s with a magnetic field magnitude H = 51.36 T, again associated with $\alpha$-$Fe_2O_3$. The doublet was fitted with $\delta$ =0.37 mm/s and $\Delta Q$ =0.53 mm/s associated to the B(16c) site in the pyrochlore structure [**14-17**].

Figures 3 b) and 4 b) display the Mössbauer spectra of the HCl washed samples. The Mössbauer spectrum of $Fe_2BiSbO_7$ was fitted considering the contribution of two doublets. The doublet 1 parameters, related with the B(16c) site in the pyrochlore structure, are $\delta$ = 0.37 mm/s and $\Delta Q$ = 0.47 mm/s, indicative of $Fe^{III}$ in low spin configuration. Doublet 2 was fitted with $\delta$ = 0.31 mm/s and $\Delta Q$ = 2.32 mm/s. The unusual eight-coordination of iron in a distorted cube in the A(16d) site of this compound is responsible for the high value of the quadrupole splitting. The obtained values of the Mössbauer hyperfine parameter indicate $Fe^{III}$ ions in a high spin configuration [**18**].

The Mössbauer spectrum for $Fe_2BiTaO_7$ sample was fitted considering a single doublet, associated to the B(16c) site in the pyrochlore structure, with $\delta$ = 0.36 mm/s and $\Delta Q$ =0.52 mm/s. The occurrence of only one doublet indicates that the iron cations do not occupancy the *A*(16c) site, in accordance with the results are consistent with the Rietveld refinement analysis. The fact that the quadrupole splitting value for the $Fe_2BiTaO_7$ compound ($\Delta Q$

=0.52 mm/s) is slightly greater than $Fe_2BiSbO_7$ (ΔQ =0.47 mm/s) suggest that the octahedral coordination around $Fe^{III}$ ions in the B(16c) site is more distorted in the tantalum pyrochlore promoted by more oxygen vacancies.

**Table III.** Mössbauer parameters (relative to α-Fe)

| Sample | Element | δ(mm/s) | ΔQ(mm/s) | H (T) | % | Γ(mm/s) | Iron Site |
|---|---|---|---|---|---|---|---|
| $Fe_2BiSbO_7$ | Doublet 1 | 0.34±0.01 | 0.48±0.02 | - | 19 | 0.16±0.03 | B(16c) |
| | Doublet 2 | 0.34±0.02 | 2.11±0.21 | - | 4 | 0.28±0.07 | A(16d) |
| | Sextet | 0.37±008 | 0.1±0.007 | 51.41±0.05 | 77 | 0.23±0.01 | α-$Fe_2O_3$ |
| $Fe_2BiTaO_7$ | Doublet 1 | 0.37±0.01 | 0.53±0.02 | - | 20 | 0.18±0.01 | B(16c) |
| | Sextet | 0.36±0.005 | 0.1±0.005 | 51.36±0.04 | 80 | 0.22±0.009 | α-$Fe_2O_3$ |
| $Fe_2BiSbO_7$ HCl | Doublet 1 | 0.37±0.02 | 0.47±0.01 | - | 84 | 0.16±0.01 | B(16c) |
| | Doublet 2 | 0.31±0.09 | 2.32±0.1 | - | 15 | 0.29±0.01 | A(16d) |
| $Fe_2BiTaO_7$ HCl | Doublet 1 | 0.36±0.01 | 0.52±0.02 | - | 100 | 0.16±0.02 | B(16c) |

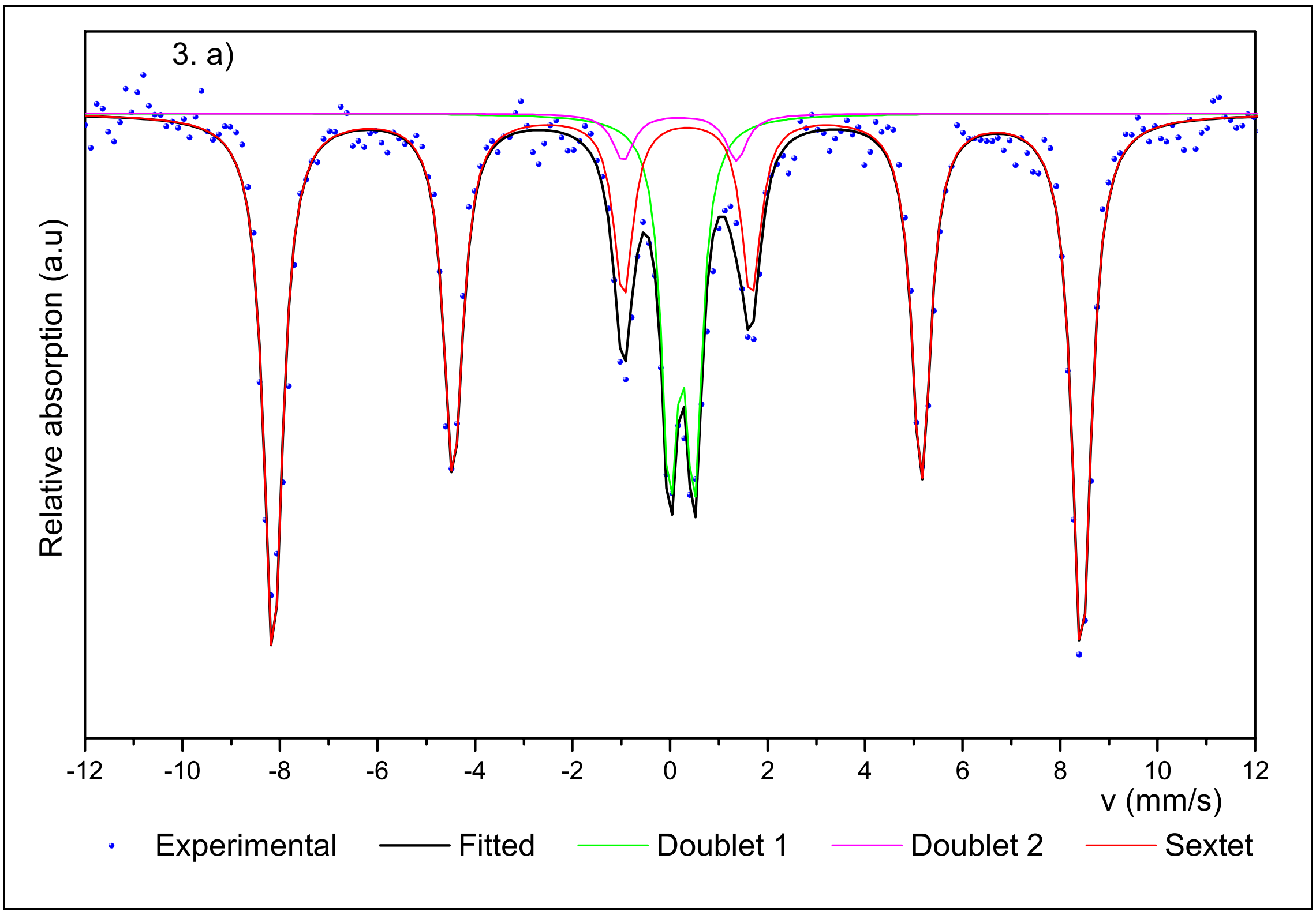

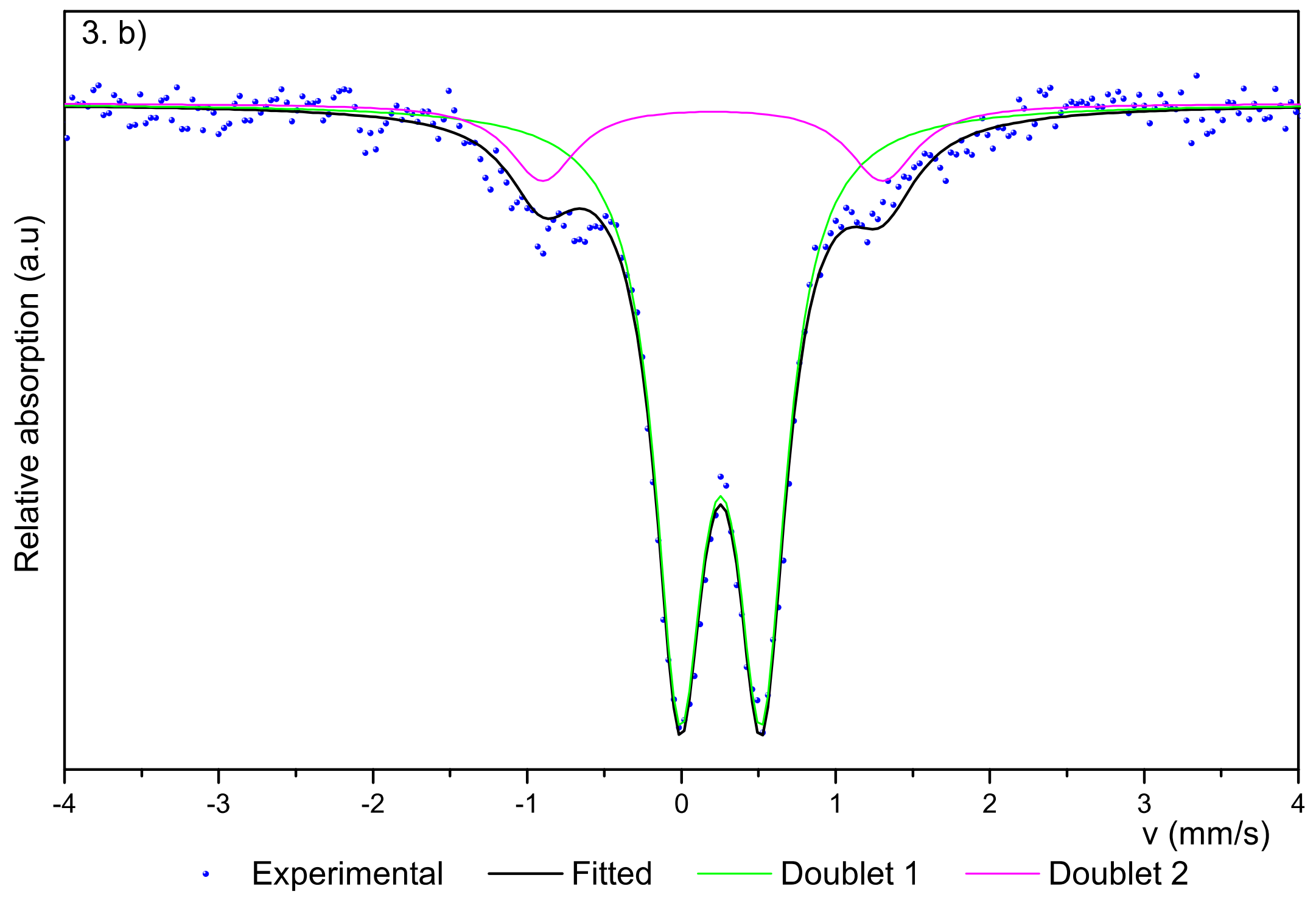

**Figure 3.** Mössbauer spectra of $Fe_2BiSbO_7$ compound before HCl treatment (3a) and after washed with HCl, corresponding to a pure pyrochlore phase (3b).

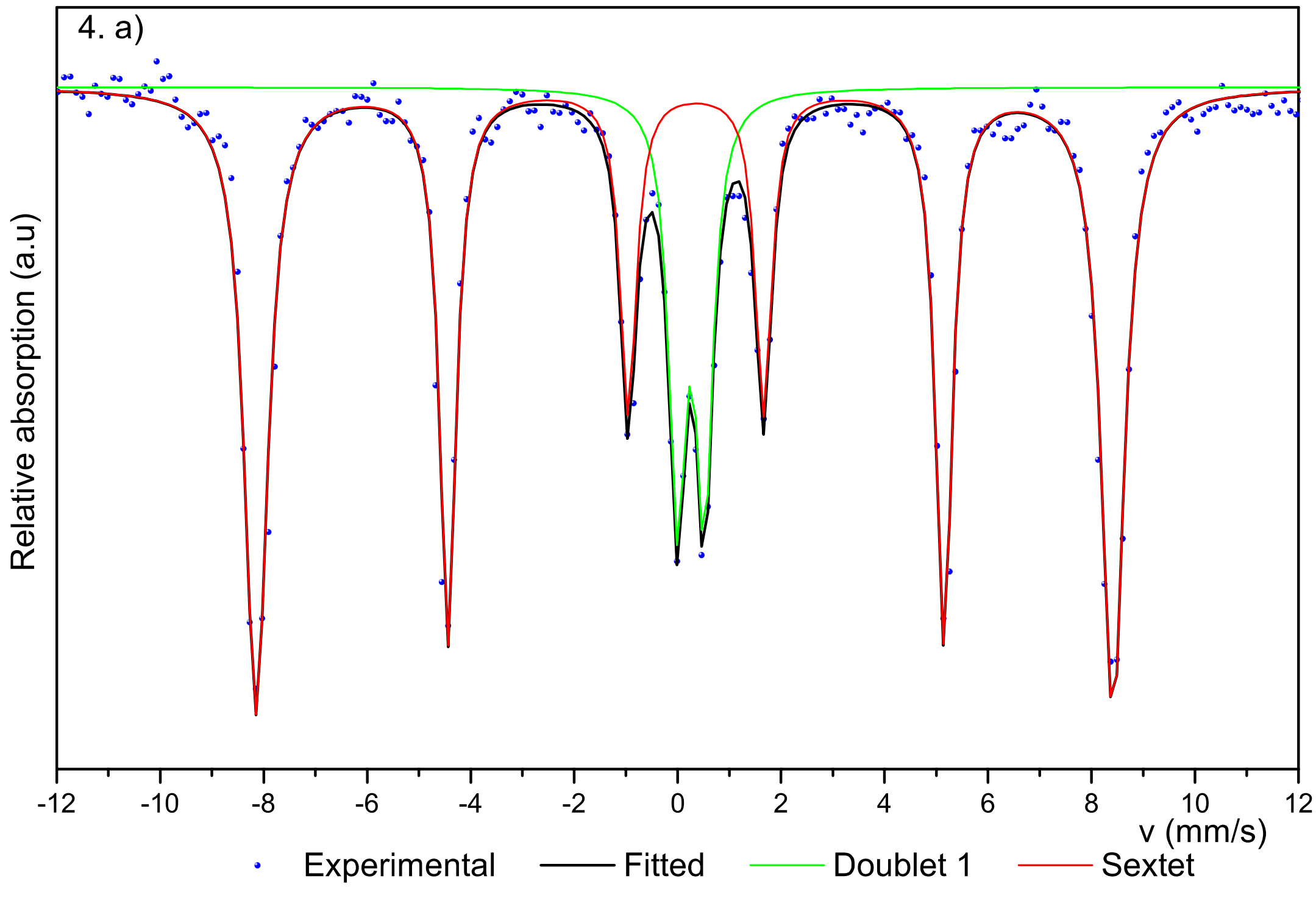

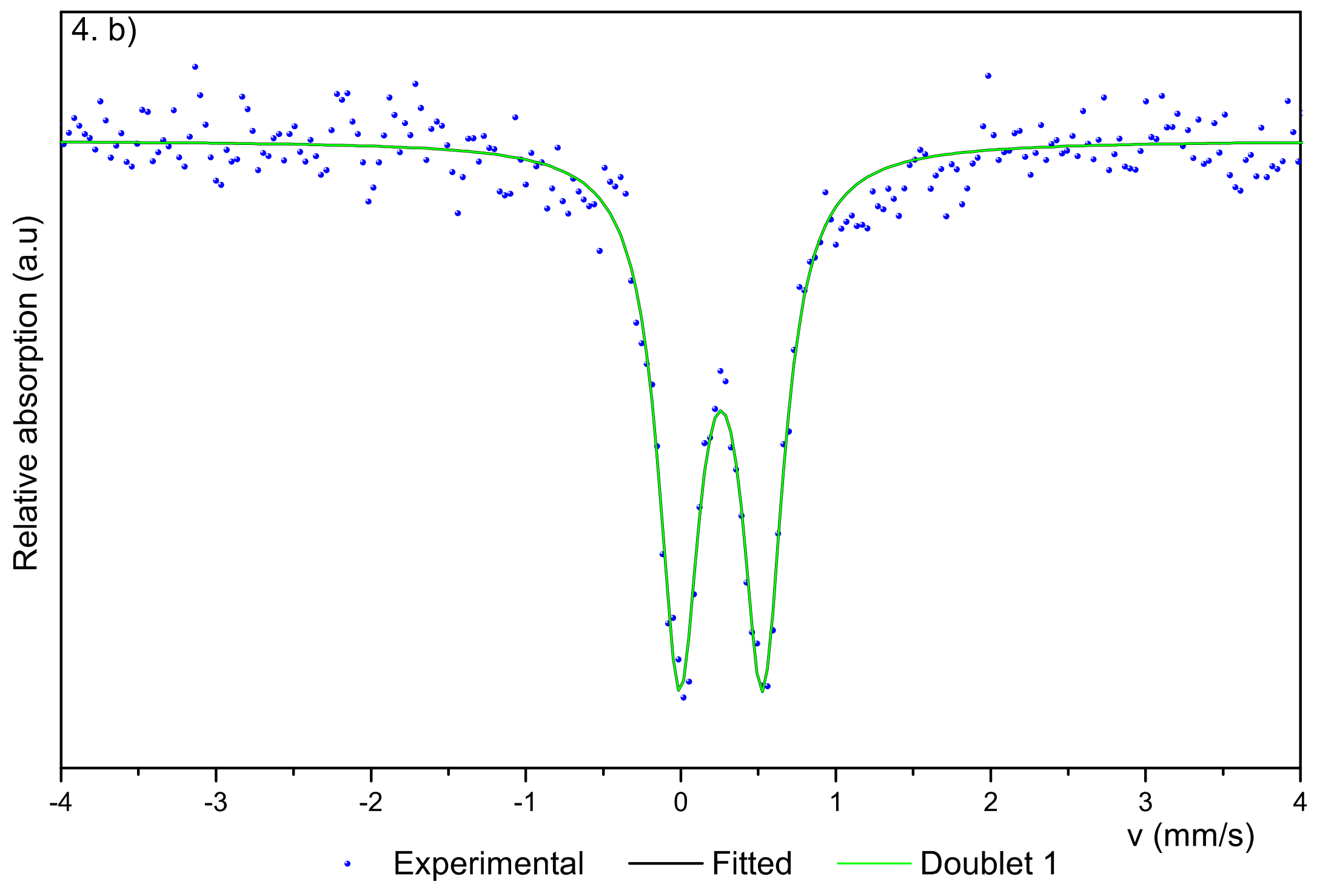

**Figure 4.** Mössbauer spectra of $Fe_2BiTaO_7$ before HCl treatment (4a) and after washed with HCl, corresponding to a pure pyrochlore phase (4b).

### 3.3 X-ray photoelectron spectroscopy

To determine the valence state of ions in the $Fe_2BiSbO_7$ and $Fe_2BiSbO_7$ pyrochlore compounds, room temperature X-ray photoelectron spectroscopy studies were performed. Figure 5 shows the X-ray photoelectron spectroscopy survey spectra from 0 to 750 eV after 30 minutes of $Ar^+$ etching the polycrystalline $Fe_2BiSbO_7$ – HCl and $Fe_2BiTaO_7$ – HCl samples. The core level energies associated with Fe, Bi, Sb, Ta and O elements are identified. Five regions related with Fe 2p, Bi 4f, Sb 4d, Ta 4f of the survey XPS spectra for both $Fe_2BiSbO_7$ and $Fe_2BiSbO_7$ compounds were analyzed. Figure 6 shows the deconvolution of the high-resolution XPS spectra for the Fe 2p core level. The XPS spectra comprise a simple spin-orbit split doublet with narrow, symmetric components in which energy split is about 13.2 eV. The detailed shape of the signal is fitted assuming the contribution of four components belonging to two different chemical states of the Fe.

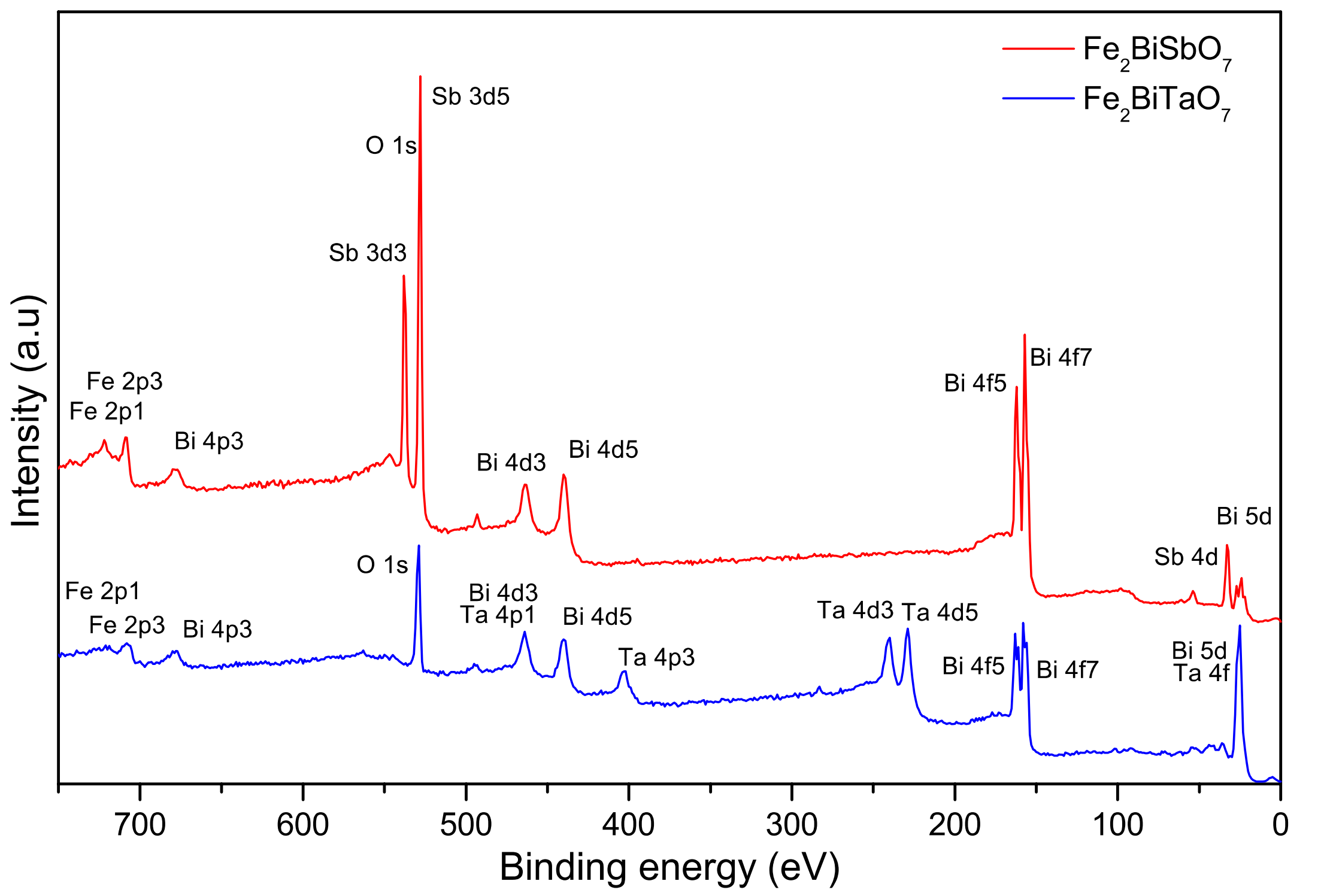

**Figure 5.** Survey X-ray photoelectron spectroscopy for the $Fe_2BiSbO_7$ (red) and Fe2BiTaO7 (blue) compounds.

A time dependent increasing rate reduction of $Fe^{3+}$ cations to $Fe^{2+}$ cations is observed as a consequence of cleaning the surface sample with $Ar^+$ bombardment in the $Fe_2BiTaO_7$ compound, in a time interval from 10 minutes to 30 minutes, as has been previously reported when Argon etching in compounds containing iron oxides is involved [**19**]. Noticeable, in the same time interval of $Ar^+$ etching there is no reduction of the $Fe^{3+}$ cations in the $Fe_2BiSbO_7$ compound. For the samples measured for the 30 minutes under $Ar^+$ etching, the Fe $2p_{1/2}$ and Fe $2p_{3/2}$ core levels binding energies are localized at 722.00 eV, 708.68 eV for the $Fe_2BiTaO_7$ compound and 721.98 eV, 708.46 eV for the $Fe_2BiSO_7$ compound, respectively. All results are in the range of the reported values for elemental $Fe^{3+}$ (NIST XPS database) [**20**].

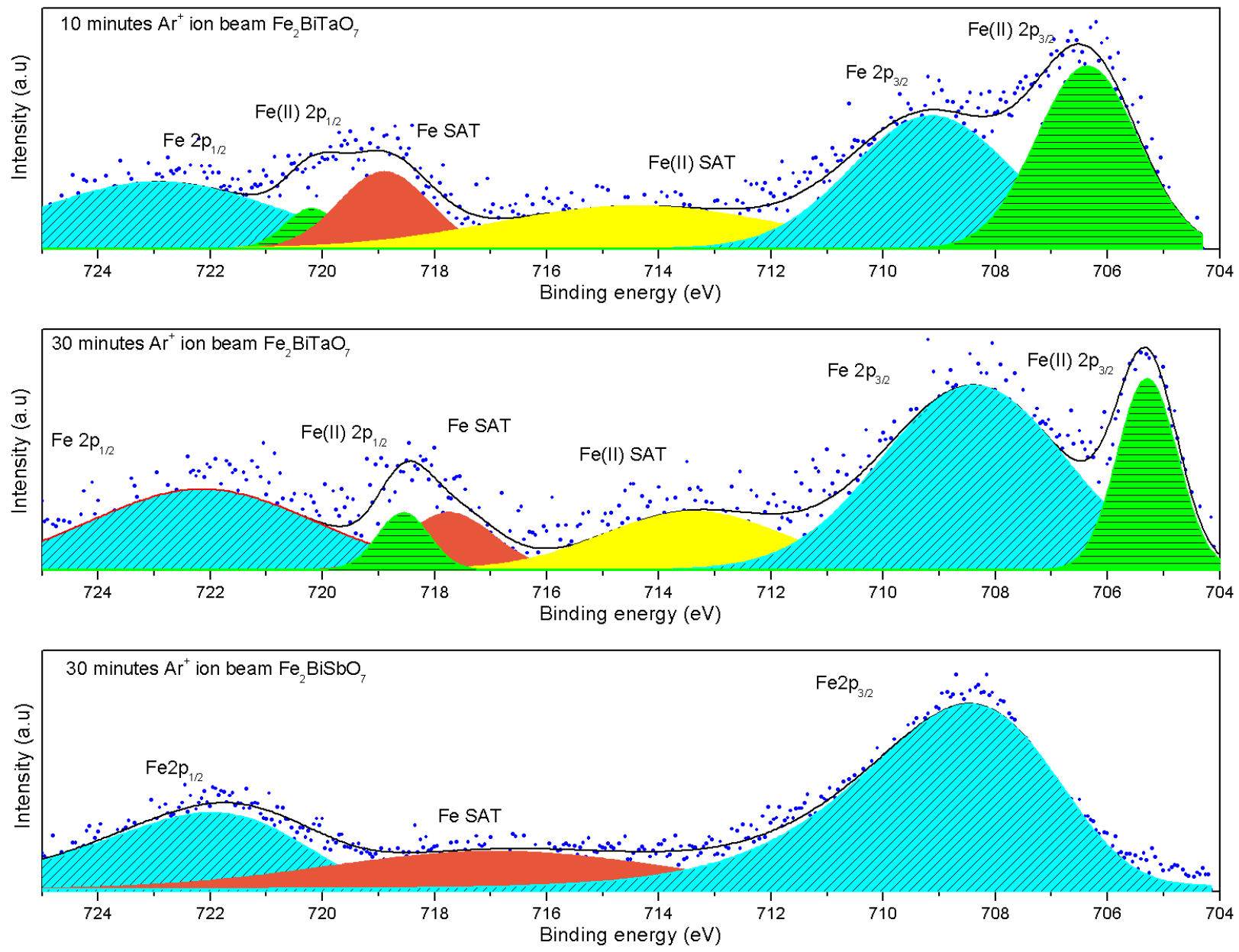

**Figure 6.** XPS spectrum at the region of iron Fe 2p. A reduction of$Fe^{3+}$ to $Fe^{2+}$, due to $Ar^{+}$ ion etching for the $Fe_2BiTaO_7$ pyrochlore, is observed.

High-resolution XPS spectra for the Bi4f core level are shown in figure 7. In the $Fe_2BiTaO_7$ compound, the features at 163.03 eV and 157.72eV correspond to the $Bi4f_{5/2}$ and $Bi4f_{7/2}$ core levels, whose energies differ by 5.31 eV [**21**], The other two features, located at 161.3 eV and 156.9 eV, correspond to the same core levels, but then in metallic Bi. For the $Fe_2BiSbO_7$ compound, the corresponding core levels energies 0.7 eV chemical shift to lowers energies: 162.33 eV and 157.02 eV, and 160.2 eV and 154.9 eV, respectively (see Table IV). The Bi4f metallic response observed is related to some bismuth reduction caused by the argon ion ($Ar^{+}$) sputtering, as reported before [**21-23**].

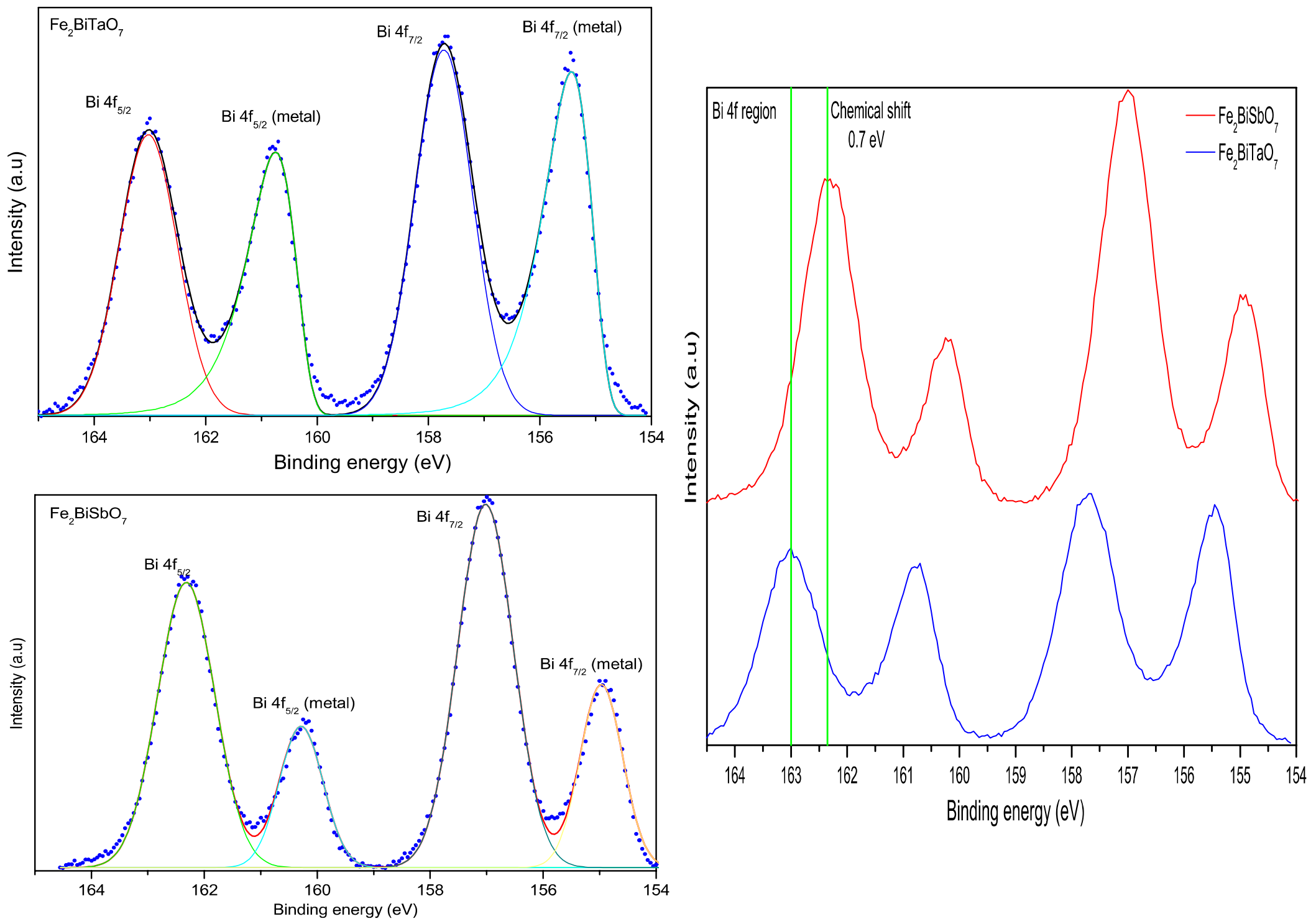

**Figure 7.** Bi 4f core level region for both compounds. A chemical shift of 0.7 eV was observed.

Figure 8a) shows, for the $Fe_2BiSbO_7$ compound, the contributions of O2s (22.04 eV), of Bi5d metal (25.01 eV, in concordance with bismuth degradation), of the $Bi5d_{3/2}$ (27.15 eV) and $Bi5d_{5/2}$ (24.06 eV) doublet, with a binding energy separation of 3.09 eV, and an interesting asymmetric feature associated with Sb4d. In order to attain a reasonable fitting for this last feature, two unresolved peaks are necessary: one for $Sb4d_{3/2}$ and the other one for $Sb4d_{5/2}$. It seems that antimony inside $Fe_2BiSbO_7$ pyrochlore structure acts with a 5+ and 3+ mixed valence [**24-26**].

Figure 8b) shows, for the $Fe_2BiTaO_7$ compound, the tantalum core level regions $Ta4f_{5/2}$ (26.67 eV) and $Ta4f_{7/2}$ (24.73 eV) with a doublet binding energy separation of 1.94 eV. Other contributions correspond to O2s (22.55 eV), $Bi5d_{3/2}$ metal (24.38 eV) and $Bi5d_{5/2}$ metal (21.51 eV), again in good agreement with bismuth degradation; structural $Bi5d_{3/2}$ (26.29 eV) and $Bi5d_{5/2}$ (23.17 eV) doublet, with a binding energy doublet separation of 3.12 eV. As can be expected, the bismuth and tantalum core level contributions overlap [**20, 27**].

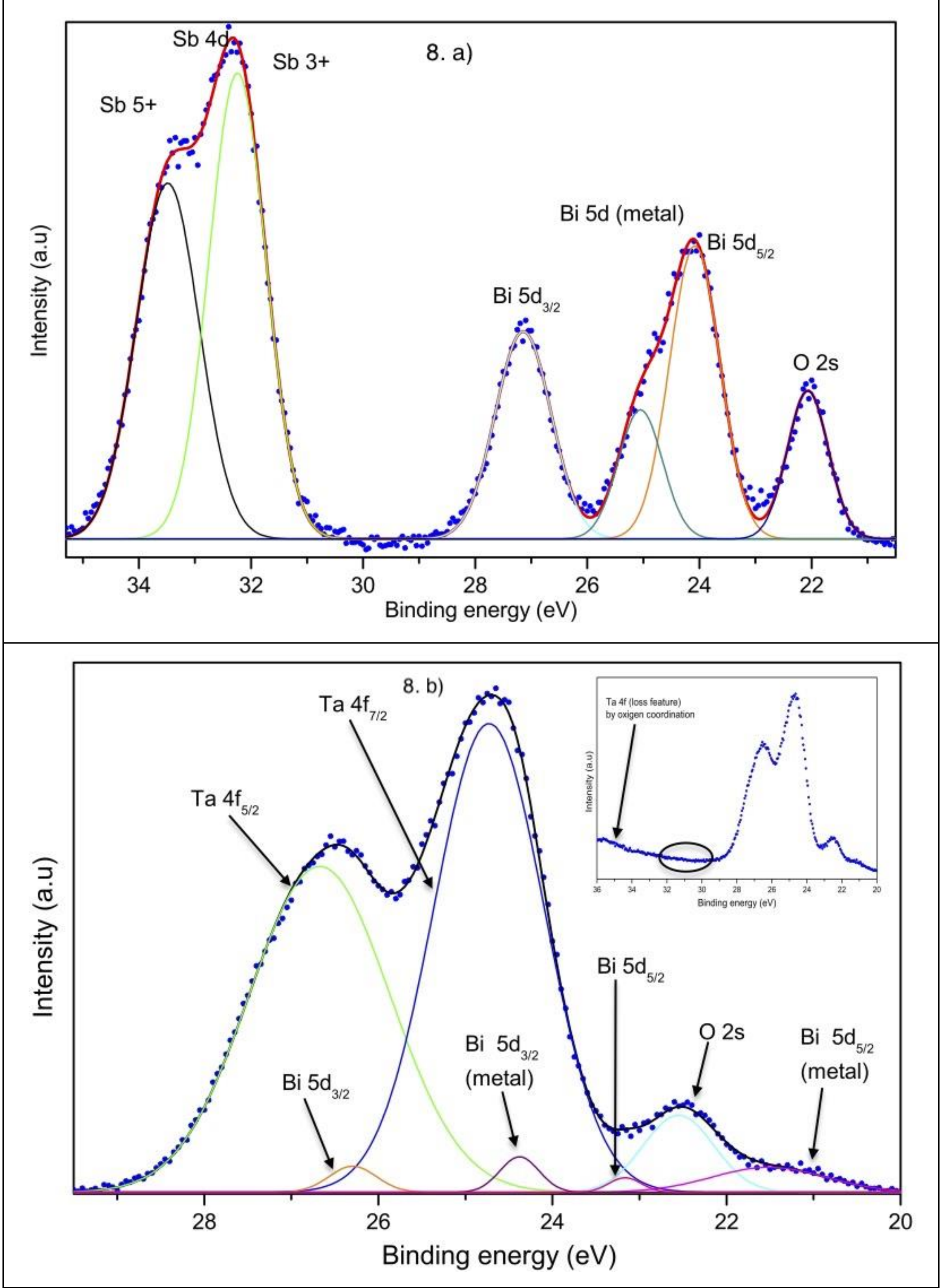

**Figure 8.** a) Sb4d core level region: a mixed valence $Sb^{5+}$-$Sb^{3+}$ state contribution is observed. b) Ta 4f region: an overlapping contribution from bismuth 5d core level is observed. Inset: the circle region shows the absence of the typically metallic tantalum loss feature region.

**Table IV.** XPS core level binding energies of both pyrochlores exposed to 30 minutes of $Ar^+$ ion bombardment.

| Core level region | BE (eV) | $\Delta E$ (eV) | BE (eV) | $\Delta E$ (eV) |
|---|---|---|---|---|
| | $Fe_2BiTaO_7$ | | $Fe_2BiSbO_7$ | |
| Fe $2p_{1/2}$ | 722.00 | 13.32 | 721.98 | 13.52 |
| Fe $2p_{3/2}$ | 708.68 | | 708.46 | |
| Fe(II) $2p_{1/2}$ | 718.57 | 13.28 | - | - |
| Fe(II) $2p_{3/2}$ | 705.29 | | | |
| Bi $4f_{5/2}$ | 163.03 | 5.31 | 162.33 | 5.31 |
| Bi $4f_{7/2}$ | 157.72 | | 157.02 | |
| Ta $4f_{5/2}$ | 26.67 | 1.94 | - | - |
| Ta $4f_{7/2}$ | 24.73 | | - | |
| Sb 4d(5+) | - | - | 33.38 | 1.20 |
| Sb 4d(3+) | - | | 32.18 | |
| O 2s | 22.55 | - | 22.04 | - |
| Bi $5d_{3/2}$ | 26.29 | 3.12 | 27.15 | 3.09 |
| Bi $5d_{5/2}$ | 23.17 | | 24.06 | |

## 4. DISCUSSION

Rietveld refinement analysis together with Mössbauer spectroscopy make clear that in the $Fe_2BiSbO_7$ compound the $Fe^{3+}$ cations can occupy both the A(16d) or the B(16c) non-equivalent crystallographic sites. In consequence, the Mössbauer spectrum (figure 3b) consists of two doublets. Both doublets have the same isomer shift δ, so their valence state is the same, but the quadrupole splitting value of the one associated with the A(16d) is about five times bigger than the one associated with the B(16d). The relative quadrupole difference between the two sites is indirectly associated with the symmetry of their surroundings: smaller values indicate higher symmetry; that is, the symmetry around the A(16d) site must be much lower than the one around the B(16c) site. The populations are 84 % for the B(16c) and only 15 % for the A(16d). The cubic geometry for 8 coordination is the less stable, so a small fraction of $Fe^{3+}$ ions in high spin configuration [**18**] in the B(16c) sites can severely distort the cube, because its smaller ionic radius, lowering its symmetry. The single doublet in the Mössbauer spectrum (figure 4b) in the $Fe_2BiTaO_7$ compound indicates that the $Fe^{3+}$ cations only occupy the B(16c) crystallographic site. Moreover, Rietveld refinement suggests the presence of $Ta^{5+}$ cations in both crystallographic sites the A(16d) and the B(16c). In such a case, the occupying facility must be associated with the relative size of competing cations.

X-ray photoelectron spectroscopy establish the oxidation states that the different cations forming both pyrochlores compounds can acquire: $Fe^{3+}$, $Bi^{3+}$, $Ta^{5+}$, $O^{2-}$, $Sb^{3+}$ - $Sb^{5+}$.

The chemical shift between $Fe_2BiTaO_7$ and $Fe_2BiSbO_7$ for similar core level regions is due to the ~ 0.04 smaller ionic radius of Ta respect to the corresponding radius of Sb, slightly affecting the core level energies values, and is not related to the oxidation state of Sb and Ta. Moreover, the isomeric shift measured in the Mössbauer spectra of the compounds is the same. Remembering that the isomer shift in Mössbauer spectroscopy is a measure of the electronic density at the nucleus (s and $p_{½}$ electrons), associated indirectly to the oxidation state, there is no discrepancy between the XPS and Mössbauer measurements.

## 5. CONCLUSION

Pyrochlore type compounds $Fe_2BiSbO_7$ and $Fe_2BiTaO_7$ where synthesized for the first time by molten salts method at 950ºC in just one hour of heat treatment. Rietveld refinement, Mössbauer spectroscopy, and XPS spectroscopy lead us to clarify the cation behavior inside the two different pyrochlore compounds. In one case, for the $Fe_2BiSbO_7$, the impossibility of the Sb cations of being eight coordinated promotes that $Fe^{3+}$ cations in high spin state partially occupying the A(16d) site in the crystal structure but at the same time this complex cation occupancy promotes a mixed oxidation state for the Sb cations in the pyrochlore structure. Meanwhile for the $Fe_2BiTaO_7$ pyrochlore the fact that $Ta^{5+}$ could be eight coordinated permits that this kind of cations partially occupies the A(16d) sites promoting a more stable crystal configuration.

## Acknowledgments

This research was partially supported by the UNAM-DGAPA-PAPIIT Program IN114416.The Authors want to thanks to Marcela Angola Bañuelos Cedano for the synthesis of a $Fe_2BiTaO_7$ sample, to M. Sc. Roberto Hinojosa Nava for his help in preparing the manuscript and to the Physicist L. Huerta for the time and support for the acquirer of the XPS data.

## REFERENCES

[1] M. Romero, R. W. Gómez, V. Marquina, J.L. Pérez-Mazariego, R. Escamilla. Physica B. 443 (2014) 90-94.
DOI: https://doi.org/10.1016/j.physb.2014.03.024

[2] Toshio Kimura, Molten salt synthesis of ceramic powders, in: Costas Sikalidis (Ed.), Advances in Ceramics – Synthesis and Characterization, Processing and Specific Applications, InTech, Croatia, ISBN 978-953-307-505-1, 2011.
DOI: 10.5772/20472

[3] Matthew L. Hand, Martin C. Stennett, Neil C. Hyatt. J. Eur. Ceram. Soc. 32 (2012) 3211-3219.
DOI: https://doi.org/10.1016/j.jeurceramsoc.2012.04.046

[4] Jingfei Luan, Zhitian Hu. Inter. J. Photoenergy 2012 (2012).
DOI: http://dx.doi.org/10.1155/2012/301954

[5] Jingfei Luan, Ningbin Guo, Biaohang Chen. Int. J. Hydrogen Energy 39 (2014) 1228-1236.
DOI: https://doi.org/10.1016/j.ijhydene.2013.11.020

[6] M. A. Subramanian, G. Aravamudan, G. V. Subba Rao. Prog. Sol. State. Chem. 15 (1983) 55-143.

[7] Maria C. Blanco, Diego G. Franco, Yamile Jalit, Elisa V. Pannunzio Miner, Graciele Berndt, Andrea Paesano Jr., Gladys Nieva, Raúl E. Carbonio. Physica B 407 (2012) 3078-3080.
DOI: https://doi.org/10.1016/j.jssc.2012.10.21

[9] P. S. Sidhu, R. J. Gilkes, R. M. Cornell, A. M. Posner, J. P. Quirk. Clays and Clay Minerals 29 (1981) 269-276.

[10] L. Lutterotti, M. Bortolotti, G. Ischia, I. Lonardelli, H.R. Wenk, Z. Krist. Suppl. 26 (2007) 125–130.

[11] Lagarec K., Rancourt, D. G., Mössbauer Spectral Analysis Software, Version 1.0, Department of Physics, University of Ottawa, 1998.

[12] http://www.xpsdata.com 2018 (accessed 02 Agost 2018)

[13] N. N. Greenwood, T. C. Gibb, *Mössbauer Spectroscopy*, Chapman and Hall Ltd London, 1971.

[14] Nicola Burriesci, Fabio Garbassi, Michele Petrera, Guido Petrini. J. Chem. Soc., Faraday Trans. 1. 3 (1982) 817-833.

[15] Frank J. Berry, John G. Holden, Michael H. Loretto, David S. Urch. J. Chem. Soc. Dalton Trans. 7 (1987) 1727-1731.

[16] F. J. Berry, A. Labarta, X. Obradors, R. Rodriguez, M. I. Sarson, J. Tejeda. Hyperfine Interactions 41 (1988) 463-466.

[17] Edilson V. Benvenutti, Luci I. Zawizlak. Polyhedron 17 (1998) 1627-1630.

[18] Ashis K. Patra, Koustubh S. Dube, Georgia C. Papaefthymiou, Jeanet Conradie, Abhik Ghosh, Todd C. Harrop. Inorg. Chem. 49 (2010) 2032-2034.
DOI: 10.1021/ic1001336

[19] M. Fondell, M. Gorgoi, M. Boman, A. Lindblad. J. Electron. Spectros. Relat. Phenomena 224 (2018) 23-26.
DOI: https://doi.org/10.1016/j.elspec.2017.09.008

[20] NIST X-ray Photoelectron spectroscopy database, Version 4.1 (National Institute of Standards and Technology, Gaithersburg, 2012); http://srdata.nist.gov/xps/, 2018 (accessed September 18, 2018 ).

[21] Kiyotaka Uchida, Akimi Ayame, Surface Science 357-358 (1996) 170-175

[22] Senol Öz, Jan Christoph Hebig , Eunhwan Jung, Trilok Singh, Ashish Lepcha, Selina Olthof, Jan Flohre, Yajun Gao, Raphael German, Paul H.M.van Loosdrecht, Klaus Meerholz, Thomas Kirchartz, SanjayMathur, Solar Energy Materials and Solar Cells, 158 (2016) 195-201.
http://dx.doi.org/10.1016/j.solmat.2016.01.035

[23] Tianyue Li, Yue Hu, Carole A. Morrison, Wenjun Wu, Hongwei Hana, Neil Robertson, Sustainable Energy Fuels, 2017,1, 308-316.
DOI: 10.1039/c6se00061d

[24] R. Reiche, J. P. Holgado, F. Yubero, J. P. Espinos, A. R. Gonzalez-Elipe. Surf. Interface Anal. 35 (2003) 256–262.
DOI: 10.1002/sia.1523

[25] Ramakanta Naik, Shuvendu Jena, R. Ganesan, N. K. Sahoo. Phys. Status Solidi B 251, No. 3, 661–668 (2014).
DOI: 10.1002/pssb.201350060

[26] V.P. Zakaznova-Herzog, S.L. Harmer, H.W. Nesbitt, G.M. Bancroft, R. Flemming, A.R. Pratt. Surface Science 600 (2006) 348–356.
DOI: 10.1016/j.susc.2005.10.034

[27] Katsuhiko Asami, Tetsuya Osaka, Tomomi Yamanobe, Ichiro Koiwa. Surf. Interface Anal. 30, 391–395 (2000).
DOI: https://doi.org/10.1002/1096-9918(200008)30:1<391::AID-SIA742>3.0.CO;2-N